Green's Functions of the Boltzmann Transport Equation with the Full Scattering Matrix for Phonon Nanoscale Transport beyond the Relaxation Time Approximation

Vazrik Chiloyan[†a], Samuel Huberman[†a,c], Zhiwei Ding[a], Jonathan Mendoza[a], Alexei A. Maznev[b], Keith A. Nelson[b], Gang Chen[a*1]

[a]*Department of Mechanical Engineering, Massachusetts Institute of Technology, Cambridge, Massachusetts 02139, USA*
[b]*Department of Chemistry, Massachusetts Institute of Technology, Cambridge, Massachusetts 02139, USA*
[c] *Department of Chemical Engineering, McGill University, Montreal, Quebec H3A 0C5, Canada*

The phonon Boltzmann transport equation (BTE) has been widely utilized to study thermal transport in solids. While for a number of materials the exact solution to the BTE has been obtained for a uniform heat flow, problems arising in micro/nanoscale heat transport have been analyzed within the relaxation time approximation (RTA). Since the RTA breaks down at temperatures low compared to the Debye temperature, this approximation prevents the study of an important class of high Debye temperature materials such as diamond, graphite, graphene and some other 2D materials. We present a full scattering matrix formalism that goes beyond the RTA approximation and obtain a Green's function solution for the linearized BTE, which leads to an explicit expression for the phonon distribution and temperature field produced by an arbitrary spatio-temporal distribution of heat sources in an unbounded medium. The presented formalism is capable of describing a wide range of phenomena, from heat dissipation by nanoscale hot spots to the propagation of second sound waves. We provide numerical results for graphene for a spatially sinusoidal heating profile and discuss the importance of using the full scattering matrix compared to the RTA.

[†] These authors contributed equally.
[*]Corresponding author: gchen2@mit.edu

## I. Introduction

Recent research on phonon-mediated thermal transport in solids demonstrated the break-down of the Fourier law of heat conduction on the micro/nanoscale when the characteristic dimension becomes comparable to the phonon mean free path. To describe non-diffusive transport that no longer obeys the ubiquitous heat equation one needs to resort to the Peierls-Boltzmann phonon transport equation (BTE) [1]. While the exact iterative solution of the BTE for the case of spatially uniform steady-state heat flow has been extensively used in the past decade for first-principles thermal conductivity calculations [2,3], micro/nanoscale heat transport involving localized heat sources and/or boundaries has been studied under the relaxation time approximation (RTA), which greatly simplifies the BTE [4]. However, the RTA fails at low temperatures (or even at room temperature for high Debye temperature materials) when umklapp phonon-phonon scattering processes are rare and normal scattering dominates. Consequently, the RTA approximation cannot be used for materials with high Debye temperature such as diamond [5] or graphene [6] and cannot capture phonon hydrodynamic phenomena such as second sound [7]. In addition, the RTA does not conserve energy (see Sec. II below), which introduces an error in the analysis that cannot be readily quantified. To assess the accuracy of the RTA in handling non-uniform and non-stationary problems, one needs to compare its results with solutions obtained with the full scattering integral.

In this paper, we aim to develop a methodology for obtaining rigorous non-stationary and non-uniform solutions of the full BTE capable of handling a wide range of problems from heat dissipation by nanoscale hot spots to the propagation of second sound waves. The iterative approach described in Ref. [8] is not easily extendable beyond the stationary and uniform special case. However, modern computational capabilities make it possible to solve the linearized BTE in the discretized wavevector space directly by matrix algebra. The diagonalization of the BTE has been previously discussed by Hardy in the context of second sound [7]; however at that time the scattering matrix could not be computed for any material. More recently, Cepellotti et al. studied the diagonalization of the scattering matrix and introduced the term "relaxons" for the corresponding eigenvectors [9,10]. However, the relaxon basis cannot be readily applied to the spatially non-uniform case as it makes the advective term in the BTE non-diagonal. Below we present a matrix-based approach for solving the BTE containing a non-uniform source term and derive the Fourier-domain Green's function with the full scattering matrix for this problem [11]. Once the Green's function is known, the temperature and phonon population distributions for an arbitrary space-time distribution of heat sources in an unbounded medium can be computed. To illustrate the range of phenomena that can be described within the proposed framework, we will present two numerical examples for graphene: (i) steady-state heat dissipation by a spatially sinusoidal heat source at room temperature, (ii) transient heat transport following impulsive spatially sinusoidal heating, yielding second sound oscillations at low temperatures.

## II. Solving the BTE with the full scattering matrix

Given an arbitrary volumetric heat generation rate $Q(\vec{r},t)$ in an infinite anisotropic crystal, we wish to calculate the phonon distribution function $f_n(r,t)$ and temperature response $T(\vec{r},t)=T_0+\Delta T(\vec{r},t)$, where $T_0$ is the background reference temperature, and $\Delta T$ is the temperature change due to the heating $Q$. Assuming that deviations from the thermal equilibrium distribution are small, the linearized phonon BTE with the full scattering matrix takes the form [12]:

$$\frac{\partial f_n}{\partial t} + \vec{\mathrm{v}}_n \cdot \vec{\nabla} f_n = Q_n \frac{N\upsilon}{\hbar\omega_n} + \sum_j W_{n,j}\left(f_j^0 - f_j\right) \quad (1)$$

where $n$ is a short-hand index for a given phonon mode (defined by a branch and wavevector in the Brillouin zone), $\omega_n$ is the phonon mode frequency, $\vec{\mathrm{v}}_n$ is the phonon group velocity, $f_n^0$ is the equilibrium distribution function, which is given by the Bose-Einstein distribution $f_n^0 = f_{BE}\left(\frac{\hbar\omega_n}{k_B T}\right)$ defined as $f_{BE}(x) \equiv \frac{1}{e^x - 1}$ , $\hbar$ is the reduced Planck constant, $k_B$ is the Boltzmann constant, and we define $N$ to be the number of discretized points in the Brillouin zone, and $b$ to be the number of phonon dispersion branches for the crystal, so that $M = bN$ is the number of phonon modes. The $d$-dimensional volume of the crystal unit cell is given by $\upsilon$ (which will be an area for a 2D material). The discrete sums can be converted to continuous integrations over the Brillouin zone as $\sum_n(\cdot) = \sum_{s=1}^{b}\sum_{\vec{k}}(\cdot) \to \sum_{s=1}^{b} N\upsilon \int \frac{d^d k}{(2\pi)^d}(\cdot)$, where $s$ is an index running over $b$ phonon branches. The matrix $W_{n,j}$ is a general scattering matrix of dimensions $M \times M$ describing the scattering rate between phonon states $n$ and $j$, acting on the difference between the equilibrium and non-equilibrium distribution functions [13]. The RTA is the simple case of a diagonal matrix $W_{n,j} = \frac{1}{\tau_n}\delta_{n,j}$, where $\tau_n$ are the relaxation times. The volumetric heat generation rate for a given mode is given by $Q_n = p_n Q$, where $Q$ is the macroscopic volumetric heat generation rate, and the values $p_n$ describe the distribution of heating among the phonon modes, which has been shown to have a large effect on the nanoscale thermal transport [14]. These values are non-negative and normalized so that $\sum_n p_n = 1$. The thermal distribution of the source is the particular case $p_n = c_n / C$, where $c_n$ is the heat capacity of a phonon mode at the reference temperature $T_0$, given by $c_n = \frac{\hbar\omega_n}{N\upsilon}\frac{\partial}{\partial T_0} f_{BE}\left(\frac{\hbar\omega_n}{k_B T_0}\right) = \frac{k_B}{N\upsilon}\left[\frac{\frac{\hbar\omega_n}{2k_B T_0}}{\sinh\left(\frac{\hbar\omega_n}{2k_B T_0}\right)}\right]^2$, and $C$ is the total volumetric heat capacity, $C = \sum_n c_n$.

The temperature $T$ is defined as the value for which the equilibrium energy density of phonons matches the nonequilibrium energy density, i.e.:

$$\frac{1}{N\upsilon}\sum_n \hbar\omega_n f_{BE}\left(\frac{\hbar\omega_n}{k_B T}\right) = \frac{1}{N\upsilon}\sum_n \hbar\omega_n f_n \tag{2}$$

If we linearize the equilibrium distribution function in terms of the temperature rise above the background, we obtain:

$$f_n^0 \approx f_{BE}\left(\frac{\hbar\omega_n}{k_B T_0}\right) + \frac{N\upsilon}{\hbar\omega_n} c_n \Delta T \tag{3}$$

To simplify, we introduce the deviational phonon energy density per mode, $g_n \equiv \frac{\hbar\omega_n}{N\upsilon}\left[f_n - f_{BE}\left(\frac{\hbar\omega_n}{k_B T_0}\right)\right]$ and $g_n^0 \equiv \frac{\hbar\omega_n}{N\upsilon}\left[f_n^0 - f_{BE}\left(\frac{\hbar\omega_n}{k_B T_0}\right)\right] = c_n \Delta T$, to yield the linearized BTE in terms of the deviational phonon energy density after proper scaling of Eq. (1):

$$\frac{\partial g_n}{\partial t} + \vec{\mathrm{v}}_n \cdot \vec{\nabla} g_n = Q p_n + \sum_j \omega_n W_{n,j} \frac{1}{\omega_j}\left(c_j \Delta T - g_j\right) \tag{4}$$

The BTE given by Eq. (4) describes transport where not only the deviation from the equilibrium distribution at the local temperature is small, but the deviation of the latter from the background constant temperature distribution is also small. The energy density above the background is given simply by $\sum_n g_n$ and the heat flux by $\sum_n g_n \vec{\mathrm{v}}_n$. In this linearized regime, the scattering matrix $W$ depends on the background temperature $T_0$ but not on the temperature rise $\Delta T$. Linearizing Eq. (2) gives the temperature rise as the ratio of the nonequilibrium energy density of phonons divided by the heat capacity:

$$\Delta T = \frac{1}{C} \sum_n g_n \tag{5}$$

The full scattering matrix must be energy conserving. This means that summing over the scattering matrix term on the right hand side of Eq. (4) must yield zero, regardless of the distribution $g_n$ [12]. If we insert the temperature of Eq. (5) into Eq. (4), and sum over all modes, energy conservation for an arbitrary distribution of modes $g_n$ will require:

$$\sum_{i,j} \omega_i W_{i,j} \frac{1}{\omega_j} \left( \frac{c_j}{C} - \delta_{j,n} \right) = 0 \tag{6}$$

which must be true for every phonon mode indexed by $n \in \left[ M \right]$ here. For a scattering matrix which satisfies the condition of Eq. (6), the system is energy conserving in any heat transfer configuration. In the RTA, the diagonal form of the scattering matrix inserted into Eq. (6) yields the following condition for the relaxation times in order for the system to be energy conserving:

$$\frac{1}{\tau_n} = \frac{1}{C} \sum_j \frac{c_j}{\tau_j} \tag{7}$$

Since the index *n* does not appear on the right-hand side of Eq. (7), the relaxation time must be the same for every mode *n*, i.e., the only energy conserving diagonal matrix *W* is an identity matrix with a single relaxation time. Consequently, the use of realistic phonon relaxation times in the RTA violates the conservation of energy. What is typically done to circumvent the violation of energy conservation in the RTA is a re-definition of the temperature, such that the energy conservation equation is used to obtain a pseudo-temperature [15,16] as opposed to the conventional temperature given by Eq. (5). The full scattering matrix is energy conserving by construction, although due to the numerical broadening used to approximate the delta functions in the matrix elements, small deviations from energy conservation may occur [13].

To solve for the phonon distribution for a system with no boundaries, we take the spatial and temporal Fourier transform of Eq. (4) to convert the differential equation into an algebraic matrix equation, and find the Fourier transform of the deviational non-equilibrium distribution function in terms of the temperature:

$$\tilde{g}_n = \tilde{Q}\sum_j A^{-1}_{n,j} p_j + \Delta\tilde{T}\left( c_n - i\sum_j A^{-1}_{n,j}\left(\omega + \vec{q}\cdot\vec{\mathrm{v}}_j\right)c_j \right) \tag{8}$$

where the matrix $A$, whose inverse appears in Eq. (8), is defined as $A_{n,j} = W_{n,j}\dfrac{\omega_n}{\omega_j} + i\delta_{n,j}\left(\omega + \vec{q}\cdot\vec{\mathrm{v}}_n\right)$

, tilde denotes the Fourier transform, $\omega$ represents the temporal frequency from the Fourier transform, not to be confused with the frequency of a phonon mode $\omega_n$, and $\vec{q}$ is the spatial wave vector from the Fourier transform. We find the temperature by inserting Eq. (8) into Eq. (5) and solving to obtain:

$$\Delta\tilde{T} = \tilde{Q}\frac{\sum_{n,j} A^{-1}_{n,j} p_j}{i\sum_{n,j} A^{-1}_{n,j} c_j\left(\omega + \vec{q}\cdot\vec{\mathrm{v}}_j\right)} \tag{9}$$

By inserting the temperature of Eq. (9) into Eq. (8) we obtain the energy density distribution for each phonon mode:

$$\tilde{g}_n = \tilde{Q}\sum_j A^{-1}_{n,j} p_j + \tilde{Q}\frac{\sum_{n,j} A^{-1}_{n,j} p_j}{i\sum_{n,j} A^{-1}_{n,j} c_j \left(\omega + \vec{q}\cdot\vec{\mathrm{v}}_j\right)}\left(c_n - i\sum_j A^{-1}_{n,j}\left(\omega + \vec{q}\cdot\vec{\mathrm{v}}_j\right)c_j\right) \tag{10}$$

Eqs. (9) and (10) constitute the main result of this work. For $Q = 1$ (i.e., for $Q$ given by a Dirac delta function in time and space), we get Fourier-domain Green's functions of the full scattering matrix BTE. For an arbitrary heat source *Q,* the temperature and phonon distribution as functions of time and position can be obtained by an inverse Fourier transform. The computational difficulty lies in inverting the large matrix *A*, whose size is determined by the discretization of the Brillouin zone and the number of the phonon branches of the crystal, at each point of the (discretized) Fourier-space.

We note that the previously obtained RTA Green's function solution of Ref. [17] is not a particular case of Eq. (9). Since Eq. (9) utilizes the definition of temperature given by Eq. (5) and an energy conserving scattering matrix *W* that satisfies Eq. (6), the temperature field obtained with Eq. (9) will not be the same as the pseudo-temperature obtained with a non-energy-conserving scattering matrix *W* in the RTA. However, if Eq. (5) is replaced by the equation for the pseudo-temperature [17], then following the same procedure as described above, we get a result for a diagonal scattering matrix that is equivalent to Eq. (9) of Ref. [17].

**III Examples**

This presented formalism allows us to study thermal transport in the micro/nanoscale regime. As the first example, we consider the one-dimensional steady state thermal grating, in which the heat source is constant in time and sinusoidal in space, $Q = \bar{Q}e^{i\vec{q}\cdot\vec{r}}$. This is a grating in one dimension along the direction of the vector $\boldsymbol{q}$ with a grating period $\lambda = 2\pi / q$. Inserting the Fourier-transform of the source function in Eq. (9), we get the temperature distribution

$$\Delta T = \bar{Q} e^{iq\cdot r} \frac{\sum_{n,j} A_{n,j}^{-1} p_j}{i \sum_{n,j} A_{n,j}^{-1} c_j \left( q \cdot \mathrm{v}_j \right)} \tag{12}$$

where the matrix $A$ is now given by the steady state form $A_{n,j} = W_{n,j} \frac{\omega_n}{\omega_j} + i\vec{q} \cdot \vec{\mathrm{v}}_n \delta_{n,j}$. It is instructive to compare this result with the temperature profile given by the heat diffusion equation in the same geometry, $\Delta T_{\mathrm{Fourier}} = \bar{Q} e^{i\vec{q}\cdot\vec{r}} \frac{1}{q^2 k_{\hat{q}}}$ where $k_{\hat{q}} \equiv \hat{q}^T K \hat{q}$ is the element of the thermal conductivity tensor in the direction of the thermal grating. The spatial temperature distributions predicted by the Fourier heat conduction equation and by the BTE are identical: both are sinusoids of the same spatial wavevector $q$ as the volumetric heating profile. However, the expression for the amplitude of the temperature modulation are different. One can define an effective thermal conductivity by matching the Fourier temperature profile to the solution of the BTE from Eq. (12). The effective thermal conductivity depends on the grating spacing $\lambda$,

$$k_{\hat{q}} = \frac{1}{q^2} \frac{\sum_{n,j} A_{n,j}^{-1} c_j i\vec{q} \cdot \vec{\mathrm{v}}_j}{\sum_{n,j} A_{n,j}^{-1} p_j} \tag{13}$$

To provide a numerical example, we calculate $k_q$ given by Eq. (13) for graphene with a natural abundance of isotopes and compare it with the RTA result. Details of the construction of the scattering matrix can be found in a previous work by Fugallo *et al.* [13], and the parameters we used can be found in the Supplemental Material. Figure 1 shows the effective thermal conductivity of graphene obtained using the full scattering matrix as well as with the RTA as a function of the grating period for the case of a thermal source distribution $p_n = c_n / C$.

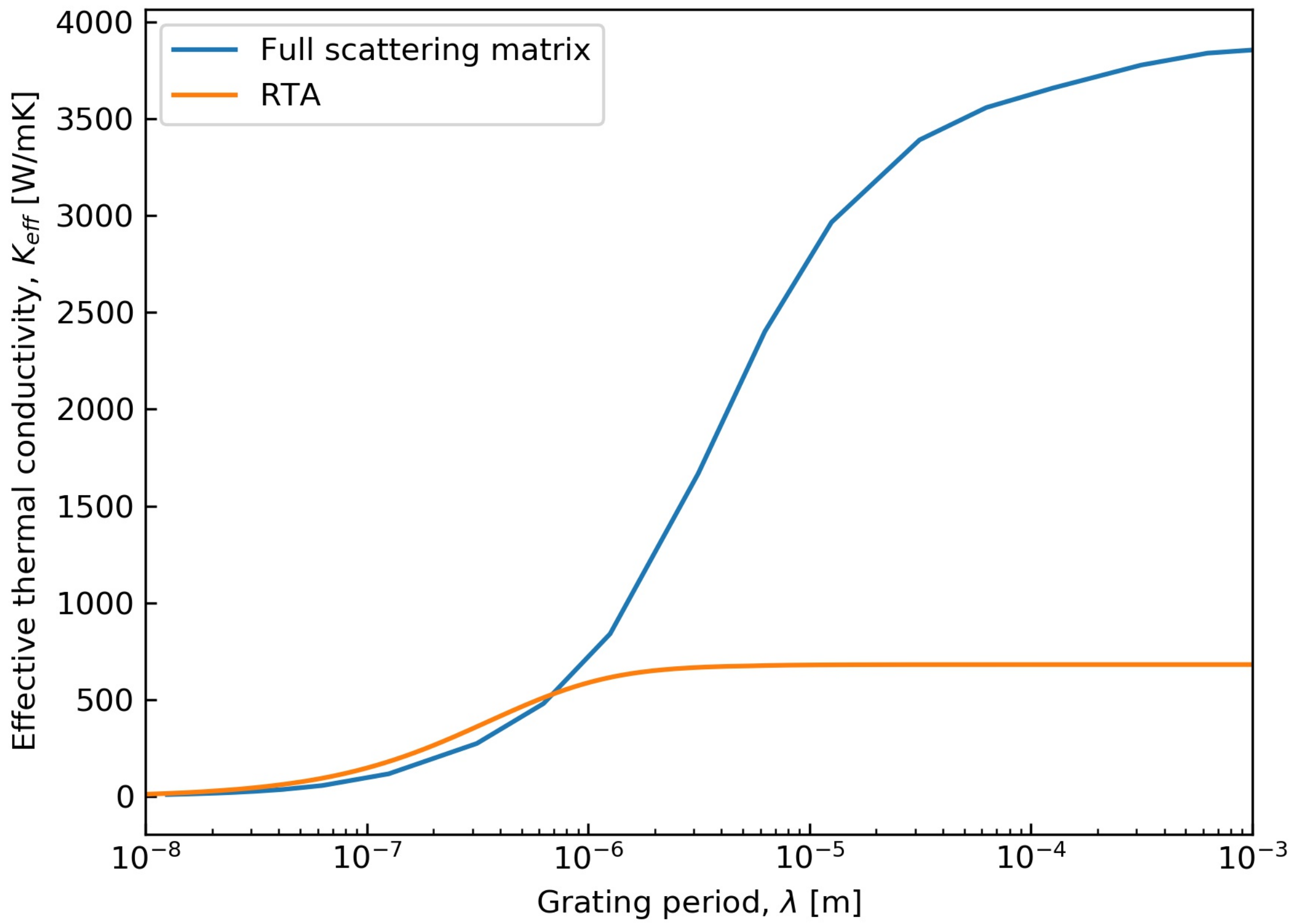

FIG. 1. Effective thermal conductivity of graphene for a steady state thermal grating at room temperature as a function of the grating period calculated with the full scattering matrix and with the RTA.

It is well known that the RTA underpredicts the macroscopic thermal conductivity of graphene [18], therefore a large discrepancy between the two curves in the limit of large $\lambda$ is not surprising. More interestingly, we find that RTA also fails to predict the size effect: the full scattering matrix calculations show that the effective conductivity decreases by 50% from the bulk macroscopic value at $\lambda \approx 4$ um, while according to the RTA, a 50 % reduction occurs at $\lambda \approx 300$ nm. We note that a localized heat source of size $L$ can be represented as a superposition of thermal gratings with periods extending down to about $2L$; thus Fig. 1 can be used to qualitatively predict the size effect in the heat dissipation from microscopic hot spots in graphene. The temperature distribution for a given profile of a localized source can be obtained by the inverse spatial Fourier transform of the solution given by Eq. (9).

Our methodology also enables modeling of transport induced by transient heat sources. As an example, we consider a transient thermal grating with the heating profile $\mathrm{Q} = \overline{\mathrm{Q}} e^{i\vec{q}\cdot\vec{r}} \delta(t)$, where a sinusoidal heat pattern is rapidly deposited into the system. Experimentally, such a source can be created by the interference two short laser pulses; the laser-based transient grating technique has been used extensively to study phonon-mediated thermal transport [19]. The Fourier transform of the temperature distribution is numerically calculated using Eq. (9), and then the inverse temporal Fourier transform yields the amplitude of the thermal grating as a function of time. Temperature responses at 100, 200 and 300 K for isotopically-pure graphene for a grating period of 10 μm are shown in Fig. 2.

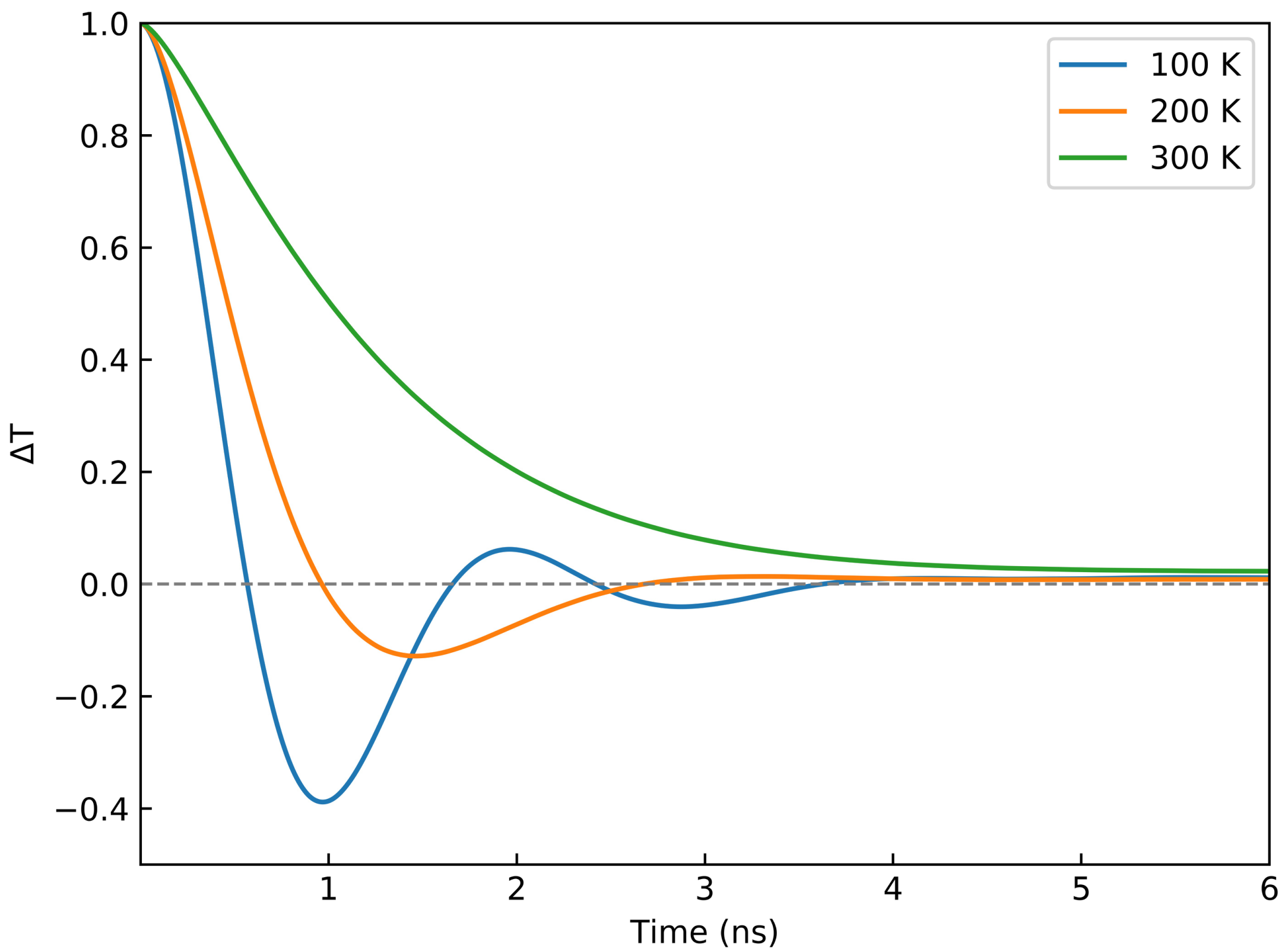

FIG 2: The amplitude of the temperature modulation in a transient thermal grating versus time for isotopically-pure graphene for a grating period of 10 μm at different background temperatures.

The time-dependent oscillations (i.e., sign changes of temperature modulation amplitude) at lower temperatures (< 200 K) indicate standing temperature waves, i.e. second sound, with a wavelength given by the transient grating spatial period λ and the speed determined by the ratio of the wavelength to the period of oscillation. These oscillations are signatures of the phonon hydrodynamic regime, where the frequency of normal scattering must be much greater than the frequency of Umklapp scattering [20]. Our approach enables one not only to predict the window of temperatures where one can expect second sound to exist [21], but simulate the temperature waves that can be experimentally observed. Indeed, recently transient grating measurements have revealed second sound in graphite at temperatures exceeding 100 K [22, 23].

## IV Conclusions

We have described a Green's function treatment of the BTE with full scattering matrix enabling calculations of the temperature and phonon population distributions produced by a heat source with an arbitrary spatio-temporal dependence. The methodology presented extends the rigorous *ab-initio* framework previously used to compute macroscopic thermal conductivity values to problems involving transient and spatially non-uniform heat flux. It allows a wide variety of thermal transport phenomena and heating geometries to be studied and will be particularly useful where both the heat equation and the RTA fail, for example, in studying nanoscale heat transport in high thermal conductivity materials and phonon hydrodynamic phenomena such as second sound.

We are grateful to Lorenzo Paulatto for his help with the construction of the scattering matrix. This work was supported by S$^3$TEC, an Energy Frontier Research Center funded by the U.S. Department of Energy, Office of Basic Energy Sciences, under Award No. DE- SC0001299/DE-FG02-09ER46577 (for thermoelectric materials) and by Multidisciplinary University Research Initiative (MURI) program, Office of Naval Research under a Grant No. N00014-16-1-2436 through the University of Texas at Austin (for high thermal conductivity materials).